\documentclass[10pt,A4,conference]{IEEEtran}
\usepackage[T1]{fontenc}
\usepackage{dsfont}
\usepackage[usenames,dvipsnames]{pstricks}
\usepackage{epsfig}
\usepackage{float}
\usepackage{enumerate}
\usepackage{balance}
\usepackage{color}
\usepackage{caption}
\usepackage{algorithm}
\usepackage[noend]{algpseudocode}
\algrenewcommand\Return{\State \algorithmicreturn{} } 
\usepackage{hyperref}
\hypersetup{
    colorlinks=true,
    linkcolor=blue,
    filecolor=magenta,      
    urlcolor=cyan,
		citecolor=blue,
}

\usepackage{amssymb}
\usepackage{amsthm}
\usepackage{array}
  \usepackage{graphicx}
  
\usepackage{epsf} 
\usepackage{psfrag}
\usepackage{multirow}
\newtheorem{theorem1}{Theorem}
\newtheorem{proposition}[theorem1]{Proposition}

\newtheorem{theorem3}{Theorem}
\newtheorem{corollary}[theorem3]{Corollary}

\usepackage{optidef}

\begin{document}
\title{Learn to Fly: A Distributed Mechanism for Joint 3D Placement and Users Association in UAVs-assisted Networks}

\author{
    \IEEEauthorblockN{Hajar El Hammouti\IEEEauthorrefmark{1}, Mustapha Benjillali\IEEEauthorrefmark{2}, Basem Shihada\IEEEauthorrefmark{1}, and Mohamed-Slim Alouini\IEEEauthorrefmark{1}}
    \IEEEauthorblockA{\IEEEauthorrefmark{1} King Abdullah University of Science and Technology
(KAUST), Thuwal, KSA.\\
	\{hajar.hammouti,basem.shihada,slim.alouini\}@kaust.edu.sa
    \IEEEauthorblockA{
	\IEEEauthorrefmark{2}	STRS Lab, National Institute of Posts and Telecommunications (INPT), Rabat, Morocco.
    }
  benjillali@ieee.org
}
}

\maketitle

\begin{abstract}
In this paper, we study the joint 3D placement of unmanned aerial vehicles (UAVs) and UAVs-users association under bandwidth limitation and quality of service constraint. In particular, in order to allow to UAVs to dynamically improve their 3D locations in a distributed fashion while maximizing the network's sum-rate, we break the underlying optimization into 3 subproblems where we separately solve the 2D UAVs positioning, the altitude optimization, and the UAVs-users association. First, given fixed 3D positions of UAVs, we propose a fully distributed matching based association that alleviates the bottlenecks of bandwidth allocation and guarantees the required quality of service. Next, to address the 2D positions of UAVs, we adopt a modified version of K-means algorithm, with a distributed implementation, where UAVs dynamically change their 2D positions in order to reach the barycenter of the cluster that is composed of the served ground users. In order to optimize the UAVs altitudes, we study a naturally defined game-theoretic version of the problem
and show that under fixed UAVs 2D coordinates, a predefined association scheme, and limited-interferences, the UAVs altitudes game is a non-cooperative potential game where the players (UAVs) can maximize the limited-interference sum-rate by only optimizing a local utility function. Therefore, we adopt the best response dynamics to reach a Nash equilibrium of the game which is also a local optimum of the social welfare function. Our simulation results show that, using the proposed approach, the network's sum rate of the studied scenario is improved by $200\%$ as compared with the trivial case where the classical version of K-means is adopted and users are assigned, at each iteration, to the closest UAV.

\end{abstract}

\section{Introduction}

As cities grow and become more developed, they rely on more technology to offer a wide range of sophisticated services and improve citizens quality of life. One of these key technologies, that is playing and will continue to play a vital role in today's and future smart cities, is the unmanned aerial vehicles (UAVs) technology, also known as \textit{drones}. In the near future, thousands of drones are expected to navigate autonomously over cities to deliver a plethora of services such as traffic reporting, package delivery and public surveillance~\cite{Survey2016}. The main virtue of such technology is the high mobility of drones, their rapid deployment, and the extremely wide range of services they can provide.  
However, deploying UAVs will pose a number of challenges. Clearly, when a drone is used to accomplish some given tasks, it is essential to design its trajectory, minimize its energy, and maximize the profit of its mission. Furthermore, in order to control the drone remotely and to communicate with ground users, it is important to study the nature of the air-to-ground channel~\cite{al2014modeling}, manage interferences~\cite{ravi2016downlink} and achieve the quality of service that satisfies the communication requirements~\cite{mozaffari2015drone}.

In this paper, we study the joint optimization of 3D positioning of UAVs and the users-UAVs association. Although a number of recent works have provided various approaches to approximately solve such an optimization problem, the majority of these works typically set up centralized algorithms to reach the best network performance. We believe that the dynamic nature of the surrounding environment and the growing size of today's networks make it extremely difficult to implement centralized schemes to achieve optimal/near-optimal solutions. Therefore, the main thrust of this paper is to design a distributed algorithm that can be implemented on UAVs in order to achieve reliable and efficient solutions by only using local information. 
 
\subsection{Related Work}

A large body of recent work is available in the literature that addresses resource allocation for UAV networks. In~\cite{shah2017distributed}, authors study small-cells-UAVs association under backhaul capacity, bandwidth constraint, and maximum number of links limitation. They present a distributed greedy approach of low complexity, to improve the users sum rate. Although not precised in~\cite{shah2017distributed}, their greedy approach is only a $\frac{1}{2}$-approximation algorithm and, thus, cannot always guarantee a good sum rate performance. In~\cite{li2017optimal}, the authors investigate the mean packet transmission delay minimization for uplink communications in a multi-layer UAVs relay network. A gradient descent approach based on Bisection method is proposed to find the optimal power and spectrum allocation in a two-layer UAVs network. Unlike the previous works where the 3D placement of UAVs is not considered, authors in~\cite{jiang2012optimization} propose an algorithm to adjust the UAV path in order to maximize a lower bound of the users sum rate over the uplink channel. Their optimization problem is solved using a line search for both time- and space-division multiple access. A single UAV placement is also investigated in~\cite{alzenad20173} where the authors objective is to maximize the number of covered users while minimizing the transmit power of drones. The UAV horizontal and vertical locations are optimized separately without any loss of global optimality. The optimal altitude is found by solving a convex decoupled optimization problem, while the optimal 2D location is achieved by finding a solution to the smallest enclosing circle problem. Authors in~\cite{zhang2018fast} study the problem of UAVs placement in order to minimize the deployment delay. The authors propose an algorithm of low complexity to maximize the coverage of a target area using heterogeneous UAVs of different coverage radii and flying speeds. The aforementioned works either consider a single UAV setup or multiple UAVs in interference-free environment. In general, optimizing the UAV placement, in isolation, is equivalent to finding the optimal 3D location that provides a good probability of line-of-sight, but at the same time, does not result in an important path loss. In the presence of interferences, an additional constraint should be considered as UAVs locations are tightly related to interferences and any inappropriate positioning of the UAV may severely affect the network performance. 

Authors in~\cite{kalantari2017user} present a heuristic particle swarm optimization algorithm to find the 3D placement of UAVs in order to maximize, under interferences, the users sum rate. In their problem formulation, the authors consider the presence of a macro base station with a large backhaul bandwidth to serve delay-sensitive users. Under this assumption, the optimal proportion of resources allocated to UAVs backhaul is determined through a decomposition process that yields in a convex optimization problem. Although the proposed algorithm provides appreciable performance, it suggests a centralized implementation which can involve a large number of signaling messages and require a high computational effort. A distributed algorithm to improve the coverage region of drones is especially considered in~\cite{farooq2018multi}. The authors assume that the positions of Internet of Things (IoT) devices are permanently changing and provide a feedback based distributed algorithm to maximize the coverage region of drones while keeping them associated in clusters. The proposed algorithm still requires a centralized information pertaining the coordinates of the clusters centers in order to reach a good network configuration.

\subsection{Contribution}
In this paper, we are interested in an urban type environment where aerial base stations are deployed to support damaged/overloaded ground base stations. Our objective is to efficiently place the UAVs in the 3D plan and associate the users with the UAVs in order to reach an optimal value of the overall sum rate of the network. Being non-convex and NP-hard, the studied problem cannot be solved using classical convex optimization methods. Therefore, we break the underlying optimization into 3 subproblems where we separately solve the 2D UAVs positioning, the altitude optimization, and the UAVs-users association. The proposed solutions are combined into a global distributed algorithm that iteratively reaches an approximate solution in only a few iterations. To summarize, our contributions can be described as follows.
\begin{enumerate}
\item In order to address the UAVs-users assignment, we propose an efficient matching association scheme that is fully distributed, that alleviates the bottlenecks of bandwidth allocation and guarantees the required quality of service.
	\item We update the UAVs 2D coordinates using a modified K-means approach that locally maximizes the sum rate of the ground users. We show through simulation results that the proposed updating rule presents a better sum rate performance when compared with the classical K-means algorithm primarily used to minimize the sum of distances.
	\item In order to optimize the UAVs altitudes, we study a naturally defined game-theoretic version of the problem and show that under fixed UAVs 2D coordinates, a predefined association scheme, and limited interferences, the UAVs altitudes game is a non-cooperative potential game where the players (UAVs) can maximize the overall sum rate by only optimizing a local utility function. To this end, we propose a distributed approach based on a local neighborhood structure to achieve an efficient altitude solution.
	 
	\item Our simulation results show that appreciable performance can be reached as compared with classical K-means and closest UAV association. 
\end{enumerate}

\subsection{Structure}
The rest of the paper is organized as follows. The next section describes the studied system model. Section~\ref{Prob} presents the general optimization problem. In Section~\ref{Prop}, the proposed approach is described. Simulation results are described in Section~\ref{Simul}. Finally, concluding remarks and possible extensions of this work are provided. 
\subsection{Notations}
Let $\textbf{M}$ and ${m}_{ij}$ denote the matrix and its $(i,j)$-th entry respectively. The set denoted by $\mathcal{S}\times \mathcal{C}$ represents the Cartesian product of $\mathcal{S}$ and $\mathcal{C}$. $\mathbb{E}_g$ is the expectation regarding random variable $g$. Vectors are denoted using boldface letters $\textbf{x}$ whereas scalars are denoted by $x$. $|\mathcal{C}|$ denotes the cardinality of the set $\mathcal{C}$. Throughout the paper, the words \textit{UAVs} and aerial base stations (\textit{ABSs}) are used interchangeably.  

\section{System Model}\label{Sys}

\subsection{Base Stations Deployment}

Consider an area $\mathcal{A}$ where the ground base stations (GBSs) form a homogeneous Poisson point process (HPPP), $\Phi_G$, of intensity $\lambda_G$. Assume that a number of GBSs is not operational or under-functioning due to a congestion (e.g. during a temporary mass event) or a malfunction (e.g. a post-disaster scenario) of the infrastructure. The overloaded/damaged base stations are modeled by an independent thinning of $\Phi_G$ with a probability $p$. In order to support the terrestrial network, a number of aerial base stations (ABSs) is deployed following a 3D HPPP, $\Phi_A$, with the same intensity as the overloaded/damaged base stations. According to Slivnyak's theorem~\cite{franccois2009stochastic}, this intensity is equal to $p\lambda_G$. 
Let $\mathcal{B}^G$ and $\mathcal{B}^A$ be realizations of $\Phi_G$ and $\Phi_A$ respectively. We denote by $(\textbf{x}^A,\textbf{y}^A,\textbf{h})$ the 3D positions matrix of all ABSs, with $(\textbf{x}^A,\textbf{y}^A)$ the 2D locations of ABSs and \textbf{h} their altitudes vector. Let $\mathcal{U}$ be the set of ground users that need to be served by the ABSs. Although not all the GBSs are overloaded/damaged, we assume, throughout the paper, that ground users are allowed to associate with ABSs only in order to avoid any additional load to the terrestrial network.


\subsection{Air-to-Ground Channel Model}
In order to capture the distortion of the signal due to obstructions, we consider the widely adopted air-to-ground channel model where the communication links are either line-of-sight (LoS) or non-line-of-sight (NLoS) with some probability that depends on both the UAV's altitude and the elevation angle between the user and the ABS. Given a UAV $j$ with an altitude $h_j$ and a user $i$ with a distance $r_{ij}$ from the projected position of the UAV on the 2D plane, the probability of LoS is given by~\cite{al2014modeling}

\begin{equation}
p^{\rm LoS}_{ij}=\frac{1}{1+\alpha {\rm exp}(\frac{180}{\pi}{\rm arctan}\frac{h_{j}}{r_{ij}}-\alpha)},
\end{equation}
where $\alpha$ and $\beta$ are environment dependent parameters. Accordingly, the path loss between UAV $j$ and user $i$, in decibel, can be written 

\begin{equation}
L^{{\rm dB}}_{ij}\!=\!\!20 {\rm log} \Bigg(\!\frac{4 \pi f_c \sqrt{r_{ij}^2+h_j^2}}{c}\!\Bigg)+ p^{\rm LoS}_{ij}\zeta_{\rm LoS}+(1-p^{\rm LoS}_{ij})\zeta_{\rm NLoS},
\end{equation}
where  the first term formulates the free space path loss that depends on the carrier frequency $f_c$ and the speed of the light $c$. Parameters $\zeta_{\rm LoS}$ and $\zeta_{\rm NLoS}$  represent the additional losses due to LoS and NLoS links respectively.

\subsection{Average spectral efficiency}

We consider downlink communication and assume that each ground/aerial base station $j$ transmits with power $P_j$. Hence, when a frame is transmitted by an ABS $j$, it is received at user $i$ with the power $P_jg_{ij}L_{ij}$, where $g_{ij}$ accounts for the multipath fading that is considered to follow an exponential distribution with mean $\mu$. The quality of the wireless link is measured in terms of SINR, $\gamma_{ij}$, defined as follows 

\begin{equation}
\gamma_{ij}=\frac{P_jg_{ij}L_{ij}}{\sigma^2+\sum\limits_{k\neq j, k \in \mathcal{B}^A \cup \mathcal{B}^G}P_kg_{ik}L_{ik}},
\end{equation}
where $\sigma^2$ represents the power of an additive Gaussian noise.  Accordingly, the average spectral efficiency received at a user $i$ from an ABS $j$, $\eta_{ij}$, can be defined using Shannon capacity bound as the following
\begin{equation}
\eta_{ij}=\mathbb{E}_\textbf{g}\big[{\rm log}_2(1+\gamma_{ij})]. 
\end{equation}

Assume each ground user $i$ has a rate request of $R_i$. Then, in order to satisfy the user's request, UAV $j$ needs to adjust the allocated bandwidth $b_{ij}$ according to the quality of the link such that 
\begin{equation}
R_{i}=b_{ij}\eta_{ij}.
\end{equation}

\section{Problem Formulation}\label{Prob}

Let $\textbf{A}=(a_{ij})$ be the users-ABSs association matrix. Our objective is to maximize the aggregate rates requested by all the ground users by optimizing, jointly, the users-ABSs association (i.e. $\textbf{A}=(a_{ij})$) and the 3D placement of ABSs (i.e. $(\textbf{x}^A,\textbf{y}^A,\textbf{h}))$ in a way that the bandwidth limitation for all ABSs is always respected and the constraint on the quality of service is not violated. Our constrained optimization problem is formulated as follows.

\begin{maxi!}
{\textbf{A},(\textbf{x}^A,\textbf{y}^A,\textbf{h})}{\sum \limits_{j \in \mathcal{B}^A}\sum \limits_{i \in \mathcal{U}} a_{ij}R_{i}\label{objective}}
{\label{GeneralOptimizationOrigin}}{}
\addConstraint{\sum_{i} a_{ij}b_{ij}\leq B_j,\quad\forall j \in \mathcal{B}^A\label{waw1}}
{}{}
\addConstraint{ \frac{a_{ij}}{\eta_{ij}}\leq \frac{1}{\eta^{\rm min}},\quad\forall (i,j) \in \mathcal{U}\times\mathcal{B}^A\label{waw11}}
{}{}
\addConstraint{ x^{\rm min}\leq x_j^A \leq x^{\rm max}\quad \forall j \in \mathcal{B}^A\label{waw12}}
{}{}
\addConstraint{ y^{\rm min}\leq y_j^A \leq y^{\rm max} \quad\forall j \in \mathcal{B}^A\label{waw13}}
{}{}
\addConstraint{  h_j \in \mathcal{H} \quad\forall j \in \mathcal{B}^A\label{waw14}}
{}{}
\addConstraint{\sum_{j} a_{ij}\leq 1,\quad\forall i \in \mathcal{U}\label{wat}}
{}{}
\addConstraint{a_{ij}\in \{0,1\},\quad\forall (i,j) \in \mathcal{U}\times\mathcal{B}^A\label{wawf3}}
{}{}
\end{maxi!}
Constraint~(\ref{waw1}) ensures that the limitation on the bandwidth resource of each UAV is respected (each UAV $j$ has a bandwidth limit $B_j$). Constraint~(\ref{waw11}) guarantees that the average spectral efficiency is  no less than a predefined threshold $\eta^{\rm min}$. Constraint~(\ref{waw12}) and~(\ref{waw13}) show that it is necessary that the ABS 2D coordinates belong to the target area. Moreover, constraint~(\ref{waw14}) ensures that the UAVs altitudes will belong to the allowed flying altitude values described in the set of discrete altitudes $\mathcal{H}$. Constraints~(\ref{wat}) and~(\ref{wawf3}) restrict the ground user to be associated, at most, with one ABS. 

In practice, problem~(\ref{GeneralOptimizationOrigin}) is not easy to solve as it involves a non-convex objective function, non-convex constraints ((\ref{waw11}) and~(\ref{wawf3})) and a non-linear constraint (in~(\ref{waw11})). Clearly, the underlying optimization problem is a mixed integer non-linear problem (MINLP) that is NP-hard. Finding an optimal solution to such a problem may involve searching over continuous 3D coordinates for all ABSs and for every possible users-ABSs association. 

In the following, we propose a distributed approach based on a local neighborhood structure to achieve an efficient global solution to the underlying optimization problem.

\section{Proposed Approach}\label{Prop}

As stated before, the problem under analysis is mathematically challenging and finding a global optimal solution cannot be achieved using classical convex optimization methods. Our objective is to compute an approximate solution, that is not necessarily optimal, but that can be reached in only a few number of iterations.
Our idea is to break up the studied problem into subproblems that are locally solvable using combined low-complexity algorithms.

\subsection{Efficient UAVs-users Matching}\label{Matching}

To deal with the target optimization, we first assume fixed 3D locations of UAVs and propose a suitable distributed mechanism for UAVs-users association. The proposed mechanism is achieved using Gale-Shapley matching~\cite{teo2001gale} where the preferences of the UAVs, on one hand, and the users on the other hand, are both based on the quality of service (i.e. the average spectral efficiency). At each step of the algorithm, each user $i$ prefers the UAV $j$ that maximizes its $\eta_{ij}$, and similarly, each UAV $j$, on its turn, prefers to serve the user $i$ that requires the lowest $b_{ij}=\frac{R_i}{\eta_{ij}}$ to maximize the number of its served users. A description of the proposed algorithm is given in \textbf{Algorithm}~\ref{MatchingAlgo}.   


\begin{algorithm}[H]
  \caption{Users-ABSs Matching
    }\label{MatchingAlgo}
     \begin{algorithmic}[1] 
	\State{\textbf{Initialization}}
	\State{For each user $i$, sort $\eta_{ij}=\frac{R_i}{b_{ij}}$ in a decreasing order such that $\eta_{ij}>\eta_{\rm min}$, and establish a list $\mathcal{L}_{i}$  \label{lineM2}}

					\State{For each ABS $j$, sort $b_{ij}=\frac{R_i}{\eta_{ij}}$ in an increasing order, and establish a list $\mathcal{L}_{j}$\label{lineM3}}

					\State{$a_{ij}=0$ for each user $i$ and ABS $j$\label{line22}}
					\Repeat
					\For{$i\in \mathcal{U}$}
					\State{ $i$ requests to connect to $j= {\rm arg max}_{k\in \mathcal{L}_i}\{\eta_{ik}\}$\label{LineRequest}}
					\If{ $i\!=\!{\rm arg min}_{\!s\in\! \mathcal{L}_j\!}\{\!b_{sj}\}$ \& $\sum\limits_{\!\substack{c\in\!\mathcal{U},\!\\c\neq i}}\!\! a_{cj}b_{cj}\!+\!b_{ij}\!\!\leq\!\!B_{j}$\label{LineAccept1}}
           \State{$a_{ij}=1$\label{LineAccept2}}
						\Else{$\sum\limits_{\substack{c\in \mathcal{U},\\c\neq i}} a_{cj}b_{cj}+b_{ij}>B_j$}
						\If{ There exists a user $s$ s.t. $b_{ij}<b_{sj}$ \& $a_{sj}=1$ \& $\sum\limits_{\substack{c\in \mathcal{U},\\c\neq i,s}} a_{cj}b_{cj}-b_{sj}+b_{ij}< B_{j}$\label{LineReject1}}
												\State{$a_{ij}=1$, $a_{sj}=0$\label{LineReject2}}
												\Else
												\State{$\mathcal{L}_{i}=\mathcal{L}_{i}\backslash \{i\}$ \& $\mathcal{L}_{j}=\mathcal{L}_{j}\backslash \{j\}$\label{Reject}}
						\EndIf
						\EndIf
						\EndFor\Until{Bandwidth limit is reached or each user has been either connected, or rejected by all its preferred ABSs. \label{Finish}}

	\end{algorithmic}
\end{algorithm}

First, each user selects the ABSs that satisfy constraint~(\ref{waw11}), and sorts them in a decreasing order by comparing their spectral efficiencies. At this step, each user has its own list of preferred ABSs (line~\ref{lineM2}). Similarly, each ABS establishes its list of preferred users by comparing the requested bandwidths (line~\ref{lineM3}). Each user sends a request to connect to its most preferred ABSs (line~\ref{LineRequest}). Each ABS accepts its most preferred users one by one until its bandwidth limit is reached and rejects the remaining users (lines~\ref{LineAccept1} and~\ref{LineAccept2}). Each rejected user attempts to connect to its second most preferred ABS, if no more bandwidth is left on this ABS, the ABS can disconnect a less desired user and replace it by the new one (lines~\ref{LineReject1} and~\ref{LineReject2})). Otherwise, the user and ABS are mutually removed from their respective preference lists (line~\ref{Reject}). The algorithm stops when all ABSs have reached their bandwidth limit or each user has been either connected, or rejected by all its preferred ABSs (line~\ref{Finish}).

\subsection{2D Placement}

At this stage of the paper, we will only deal with the 2D placement of UAVs. In particular, we assume that the users-ABSs association scheme is the one described in Subsection~\ref{Matching} and that the altitudes for all ABSs are fixed at some random values. The UAVs altitudes are addressed separately in Subsection~\ref{Alt}. Our objective is to move the UAVs towards their served ground users in the 2D plan, in sequential steps, so that the quality of the link for each group is improved, and eventually, more bandwidth is left to serve additional users. 

To this end, we propose a modified version of $K$-means algorithm~\cite{bottou1995convergence} (with $K=|\mathcal{B}^A|$) that operates in a distributed and  asynchronous (sequential) fashion. This modified version positions the UAVs as the barycentre of the \textit{served} UAVs instead of the barycentre of the \textit{closest} users as it is the case for the classical $K$-means algorithm. The procedure of the UAVs 2D placement via the modified version of $K$-means is presented in \textbf{Algorithm}~\ref{Kmeans}.


\begin{algorithm}
  \caption{2D Placement Optimization
    }\label{Kmeans}
     \begin{algorithmic}[1] 
					\State{\textbf{Initialization}}
					\State{For each ABS $j$, $(x_j^A(0),y_j^A(0))$ are chosen randomly within the target area $\mathcal{A}$\label{line1K}}
 \State{For each ABS $j$, $\mathcal{C}_j=\varnothing$\label{line2K}}
					\Repeat
					\For{$j$ in $\mathcal{B}^A$}
					\If{$j$ is active \label{lineActK}}
					\For{$i$ in $\mathcal{U}$}
					\State{Update $\eta_{ij}$\label{line3K}}
					\State{Update $\textbf{A}$ according to \textbf{Algorithm}~\ref{MatchingAlgo}\label{line33K}}

					\If{$a_{ij}=1$\label{line4K}}
					\State{$\mathcal{C}_j=\mathcal{C}_j\cup\{i\}$\label{line5K}}
					\EndIf
					\EndFor
					\State{$x_j^A(t+1)\leftarrow \frac{\sum\limits_{i \in \mathcal{C}_j} x_{i}}{|\mathcal{C}_j|}$\label{line6K}}
					\State{$y_j^A(t+1)\leftarrow \frac{\sum\limits_{i \in \mathcal{C}_j} y_{i}}{|\mathcal{C}_j|}$\label{line7K}}
					\EndIf
					\EndFor
          \Until{UAVs cannot improve their 2D locations or number of iterations has reached a predefined threshold.\label{line8K}}
	\end{algorithmic}
\end{algorithm}
Given $K$ initial positions of ABSs $(\textbf{x}^A(0),\textbf{y}^A(0))$ (line~\ref{line1K}), the algorithm groups the users with their serving ABSs determined using the association scheme described in \textbf{Algorithm}~\ref{MatchingAlgo} (line~\ref{line33K}). Accordingly, each UAV's 2D position is updated as a barycentre of its cluster (lines~\ref{line6K} and~\ref{line7K}). 
 The algorithm is implemented on board of each ABS in an asynchronous way where each ABS updates its state depending on its own clock (line~\ref{lineActK}). Each UAV has two states: active and dormant. In the active state, the ABS updates its 2D coordinates while in a dormant state, the UAV sleeps in order to save its energy and reduce exchanged signaling messages. When the position of a UAV is updated, the feedbacks of the users are updated as well: the UAV serving each user may change. This process is then repeated until none of the UAVs 2D locations are updated or the number of iterations reaches a predefined threshold (line~\ref{line8K})\footnote{The convergence of such a process to a stable equilibrium is left as part of future work}.



%

\subsection{Altitude Optimization}\label{Alt}
In this subsection, we optimize the UAVs altitude given fixed 2D coordinates of UAVs and a predefined association scheme, specifically, the one described in Subsection~\ref{Matching}.
\subsubsection{Definitions}
Throughout this section, we adopt the following definitions.
\begin{itemize}
	\item\textbf{ Neighborhood}: two base stations (regardless the fact of being ground or aerial base stations) are considered neighbors if there exists at least one user that is covered by both base stations when they are at their maximum coverage. In mathematical words, the neighborhood of a base station $j$ can be defined as follows.
	\begin{align}
	\mathcal{N}_j(\!\tau\!)\!&=\!\{k \!\in \!\mathcal{B}^A\!\cup \! \mathcal{B}^G\!, \exists i \in\mathcal{U} \text{ s.t. } \! {\rm max}_{h_j}P_j L_{ij} >\tau \nonumber \\& \quad \text{ and } {\rm max}_{h_k} P_k L_{ik}\! > \tau\},
	\end{align}
	where $\tau$ is the received signal threshold. Note that such a threshold is defined on the received power averaged over small-scale (multipath) fading. For ease of notation, we will remove the 'dependency' on $\tau$ in the rest of the paper, and note $\mathcal{N}_{j}$ instead of $\mathcal{N}_j(\tau)$.
	\item \textbf{Local sum rate function}: is the function that computes the sum rate over a local neighborhood set. Thus, instead of considering the social welfare of all base stations with all interferences, only rates from neighboring base stations  with limited interferences (coming from neighbors) are considered. Accordingly, for each ABS $j$, the local sum rate is given by 
	\begin{equation}
	U_j(\textbf{h}\!)\!=\!\!\!\!\sum\limits_{l\in \mathcal{N}_j}\!\sum\limits_{i\in\mathcal{U}} \! a_{il}b_{il} \mathbb{E}\!\Big(\!{\rm log}_2\big(\!1+\frac{P_lg_{il}L_{il}}{\sigma^2+\!\!\sum\limits_{\substack{k \in \mathcal{N}_{j},\\k\neq l}}\!\!P_k g_{ik}L_{ik}}\!\big)\!\Big).
	\end{equation}
	Note that when $\tau=0$ the local sum rate function coincides with the social welfare provided by the global objective function in equation~(\ref{objective}).
	\item \textbf{Nash equilibrium (NE):}
	A strategy profile $\textbf{h}$ is a Nash equilibrium of a game $\mathcal{G}$ if for each player $j$, $\forall h_j\neq h^*_j$~\cite{maskin1999nash}
\begin{equation}
	U_j(h_j^*,\textbf{h}_{-j}^*)\geq U_j(h_j,\textbf{h}^*_{-j}),
\end{equation}		
where $\textbf{h}_{-j}$ refers to the altitudes vector of UAVs other than $j$.
		At a Nash equilibrium no player has the incentive to unilaterally change its strategy. 
	
	\item \textbf{Potential game}: In game theory, an interesting class of games called \textit{potential} games has a specific property: the NE is a local optimum of the social welfare function also called a \textit{potential function}. Let $\mathcal{X}$ be a set of strategy profiles of a game $\mathcal{G}$. 
$\mathcal{G}$ is a potential game if there exists a potential function $F: \mathcal{X}\longrightarrow \mathbb{R}$ such that for each player ${j}$, $\forall (h_j,\textbf{h}_{-j}) \text{ and } (h_j',\textbf{h}_{-j})~\in~\mathcal{X}$~\cite{monderer1996potential} 
\begin{equation}
F(\! h_j,\textbf{h}_{-j}\!)\!-\!F(h_j'\!,\textbf{h}_{-j}\!)\!=\!U_{j}(h_j\!,\textbf{h}_{-j}\!)\!-\!U_j(h_j'\!,\textbf{h}_{-j}\!).
\end{equation}

\end{itemize}

\subsubsection{Altitudes adjustment }

Let $F(\textbf{h})$ be the sum rate of all users where only interferences from neighboring base stations are considered. This function is given by 
\begin{equation}\label{EquationF}
F(\textbf{h})=\!\!\!\!\sum\limits_{j\in \mathcal{B}^A}\!\sum\limits_{i\in\mathcal{U}} \! a_{ij}b_{ij} \mathbb{E}\!\Big(\!{\rm log}_2\big(\!1+\frac{P_lg_{ij}L_{ij}}{\sigma^2+\!\!\sum\limits_{\substack{k \in \mathcal{N}_{j},\\k\neq j}}\!\!P_k g_{ik}L_{ik}}\!\big)\!\Big).
\end{equation}
In order to account for the neighborhood and altitudes in the average spectral efficiency, we set the following notation
\begin{equation}
 \eta_{ij}^{\mathcal{N}_j}(\! h_j,\textbf{h}_{-j}\!)=\mathbb{E}\!\Big(\!{\rm log}_2\big(\!1+\frac{P_lg_{ij}L_{ij}}{\sigma^2+\!\!\sum\limits_{\substack{k \in \mathcal{N}_{j},\\k\neq j}}\!\!P_k g_{ik}L_{ik}}\!\big)\!\Big),
\end{equation}
where $L_{ij}$ is the path loss when UAV $j$ is at altitude $h_j$. Hence, when a UAV $j$ changes its altitude given fixed altitudes of its opponents, the difference in the limited-interference sum-rate can be written
\begin{flalign}
&F(\! h_j,\textbf{h}_{-j}\!)-\! F(\! h_j'\!,\textbf{h}_{-j}\!)= \!\!\!\!\sum\limits_{\substack{l\in \mathcal{B}^A\backslash \mathcal{N}_j}}\!\sum\limits_{i\in\mathcal{U}} \! a_{il}b_{il} \eta_{il}^{\mathcal{N}_l}(\! h_j,\textbf{h}_{-j}\!)+\nonumber\\&\sum\limits_{l\in \mathcal{N}_j}\!\sum\limits_{i\in\mathcal{U}} \! a_{il}b_{il} \eta_{il}^{\mathcal{N}_l}(\! h_j,\textbf{h}_{-j}\!)-\!\!\!\!\sum\limits_{\substack{l\in \mathcal{B}^A\backslash \mathcal{N}_j}}\!\sum\limits_{i\in\mathcal{U}} \! a_{il}b_{il} \eta_{il}^{\mathcal{N}_l}(\! h^{'}_j,\textbf{h}_{-j}\!)-\nonumber\\&\sum\limits_{l\in \mathcal{N}_j}\!\sum\limits_{i\in\mathcal{U}} \! a_{il}b_{il} \eta_{il}^{\mathcal{N}_l}(\! h^{'}_j,\textbf{h}_{-j}\!).
\end{flalign}

Notice that the term $ \!\!\!\!\sum\limits_{\substack{l\in \mathcal{B}^A\backslash \mathcal{N}_j}}\!\sum\limits_{i\in\mathcal{U}} \! a_{il}b_{il} \eta_{il}^{\mathcal{N}_l}(\! h_j,\textbf{h}_{-j}\!)$ is independent of $(\! h_j,\textbf{h}_{-j}\!)$ as it does not involve UAV $j$ neighborhood. Therefore,
\begin{flalign}
&F(\! h_j,\textbf{h}_{-j}\!)-\! F(\! h_j'\!,\textbf{h}_{-j}\!)\nonumber\\&= \sum\limits_{l\in \mathcal{N}_j}\!\sum\limits_{i\in\mathcal{U}} \! a_{il}b_{il} \eta_{il}^{\mathcal{N}_l}(\! h_j,\textbf{h}_{-j}\!)-\sum\limits_{l\in \mathcal{N}_j}\!\sum\limits_{i\in\mathcal{U}} \! a_{il}b_{il} \eta_{il}^{\mathcal{N}_l}(\! h^{'}_j,\textbf{h}_{-j}\!)\nonumber\\&=\!U_{j}(h_j\!,\textbf{h}_{-j}\!)\!-\!U_j(h_j'\!,\textbf{h}_{-j}\!).
\end{flalign}

The following \textbf{Proposition} arises from the previous analysis.
\begin{proposition}\label{Propo}
Let $\mathcal{G}$ be the game where the UAVs are considered as players and the altitudes are their playing strategies. The game $\mathcal{G}$ is a potential game where the function $F$ defined by equation~(\ref{EquationF}) is the potential function.
\end{proposition}
The following result is an immediate consequence of \textbf{Proposition}~\ref{Propo}~\cite{monderer1996potential}.

\begin{corollary}\label{cor}
In a potential game, a global optimum of the potential function is a Nash equilibrium. Moreover, any Nash equilibrium is a local optimum.
\end{corollary}

Accordingly, in order to reach a local optimum of the limited-interference sum rate $F$, we can only target a NE. To this end, we adopt \textbf{Algorithm}~\ref{BestResponse}, based on best-response dynamics, to help UAVs to adaptively learn how to play a NE over iterations~\cite{swenson2018best}.

\begin{algorithm}
  \caption{Best-Response Dynamics for Altitudes Adjustment
    }\label{BestResponse}
     \begin{algorithmic}[1] 
					\State{\textbf{Initialization}}
					\State{Let $(\textbf{x}^A,\textbf{y}^A)$ be the 2D locations vector obtained using \textbf{Algorithm}~\ref{Kmeans} \label{line1B}}
					\State{For each UAV $j$, determine its neighborhood \label{line11B}}
					\Repeat
					\For{$j \in \mathcal{U}$}
					\State{$h_j^*={\rm arg max}_{h\in \mathcal{H}} U_j(h,h_j)$\label{line2B}}
					\State{Update $\eta_{ik}$ for all neighbors of $k \in \mathcal{N}_j$\label{line3B}}
					\State{Update $\textbf{A}$ using \textbf{Algorithm}~\ref{MatchingAlgo}\label{line3B}}
					\EndFor
					\Until{A NE is reached. \label{line4B}}
	\end{algorithmic}
\end{algorithm}

Assume fixed 2D locations of UAVs (line~\ref{line1B}), each UAV maximizes its utility $U_j$ over a set of discrete altitude's values $\mathcal{H}$ given fixed altitudes of other ABSs (line~\ref{line2B}). Subsequent changes are therefore fed back to the neighbors resulting in updates of the association matrix using \textbf{Algorithm}~\ref{MatchingAlgo}. This process is repeated until convergence to a NE\footnote{Convergence of best response dynamics to a NE has been proved in many works, e.g.~\cite{swenson2018best}.} (line~\ref{line4B}). Such process results in a local optimum of $F$ given fixed 2D positions of the UAVs and the predefined association scheme. 

\section{Simulation Results}\label{Simul}
\subsection{Simulation Setup}
In order to study the performance of the proposed approach, we consider $1 {\rm km}\times 1 {\rm km}$ area where a number of GBSs with intensity $\lambda_G=0.22*10^{-4}$ are randomly scattered. Assume $f_c=2$ GHz, $P_j=10 dBm$ for all base stations. In order to compute the average spectral efficiency, we use Monte Carlo simulations with $5000$ runs. The simulation settings are summarized in TABLE.~\ref{Tab}.
\begin{table}[H]
\begin{center}
\begin{tabular}{|c|c|c|c|}
\hline
Parameter & Value & Parameter & Value\\
\hline
$R_i$& Random from $[90,100]$ Mbps&
$\alpha$& $9.61$\\
\hline
$\beta$& $0.16$&
 $c$&$3.10^8m\!/s$\\
\hline
$\zeta_{\rm LoS}$& $1$ dB&$\zeta_{\rm NLoS}$ & $20$dB
\\ 
\hline
$\eta^{\rm min}$& -20 dB &
$\tau$&  -69 dBm\\
\hline
$\mu$& 1&
$p$& 0.35\\
\hline
$\lambda_G$& $0.22*10^{-4}$ GBS per $m^2$&
$|\mathcal{U}|$&$46$\\
\hline
$\mathcal{H}$& $\{\!40\!,100\!,160\!,220\!,280\!,340\!\}m$ &$\sigma^2$&-100 dBm\\
\hline
\multirow{1}{*}{UAVs } &
  \multicolumn{3}{c|}{$\{756, 696, 567, 737, 968, 631, 814,$}\\
	
	\multirow{1}{*}{bandwidth} &
	  \multicolumn{3}{c|}{$ 573, 930, 796, 742, 767,  712\}$ MHz} \\
\hline
\end{tabular}
\caption{Simulation Settings.}
\label{Tab}
\end{center}
\end{table}

\subsection{Results}

\begin{figure}[H]
\centering
		\psfrag{UAV Start Point}[][][0.65]{UAV Start Point}
		\psfrag{UAV Final Point}[][][0.65]{UAV Final Point}
		\psfrag{User}[][][0.65]{User}
		\psfrag{UAV Mouvement}[][][0.65]{UAV Movement}
    \psfrag{User-UAV association}[][][0.65]{User-UAV Association}

		\psfrag{x (m)}[][][0.7]{$x (m)$}
		\psfrag{y (m)}[][][0.7]{$y (m)$}
		 \psfrag{4.466}[][][0.65]{$4.466$}
	 \psfrag{4.468}[][][0.65]{$4.468$}
	 \psfrag{4.47}[][][0.65]{$4.47$}
	 \psfrag{4.472}[][][0.65]{$4.472$}
   \psfrag{4.474}[][][0.65]{$4.474$}
	 \psfrag{4.476}[][][0.65]{$4.476$}
	 \psfrag{4.478}[][][0.65]{$4.478$}
	
	 \psfrag{5.4118}[][][0.65]{$5.4118$}
	 \psfrag{5.412}[][][0.65]{$5.412$}
	 \psfrag{5.4122}[][][0.65]{$5.4122$}
	 \psfrag{5.4124}[][][0.65]{$5.4124$}
   \psfrag{5.4126}[][][0.65]{$5.4126$}
	 \psfrag{5.4128}[][][0.65]{$5.4128$}
	 \psfrag{5.413}[][][0.65]{$5.413$}
		\includegraphics[width=8cm,height=5cm]{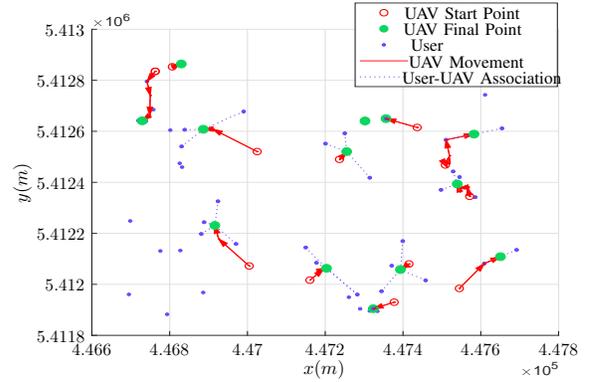}
		\caption{Movement of UAVs in 2D plan.}
	\label{Association2D}
\end{figure}

Fig.~\ref{Association2D} plots the positions of UAVs in the 2D plan. As depicted in the figure, ABSs dynamically change their 2D positions. Starting from their initial points (dots in red), the UAVs move towards the served users in a few steps before reaching their final destination (dots in green) considered as the barycenter of their served users. Clearly, due to the bandwidth limitation and the quality of service constraint, some users are left without connectivity. In the studied scenario, $30$ users from a total of $46$ uniformly distributed users are finally connected to the UAVs. It is important to note that the K-means algorithm convergence depends on the initial values of the network configuration: different initialization vectors would lead to different final results. Some initialization procedures can however be used to improve the convergence results of K-means~\cite{bottou1995convergence}.

\vspace{-0.5cm}

\begin{figure}[H]
	\centering
	 \psfrag{x (m)}[][][0.7]{$x (m)$}
	 \psfrag{y (m)}[][][0.7]{$y (m)$}
	 \psfrag{z (m)}[][][0.7]{$h (m)$}
	 \psfrag{4.465}[][][0.65]{$4.465$}
	 \psfrag{4.468}[][][0.65]{$4.468$}
	 \psfrag{4.47}[][][0.65]{$4.47$}
	 \psfrag{4.475}[][][0.65]{$4.477$}
   \psfrag{4.474}[][][0.65]{$4.474$}
	 \psfrag{4.476}[][][0.65]{$4.476$}
	 \psfrag{4.478}[][][0.65]{$4.478$}
	
	 \psfrag{5.4118}[][][0.65]{$5.4118$}
	 \psfrag{5.412}[][][0.65]{$5.412$}
	 \psfrag{5.4122}[][][0.65]{$5.4122$}
	 \psfrag{5.4124}[][][0.65]{$5.4124$}
   \psfrag{5.4126}[][][0.65]{$5.4126$}
	 \psfrag{5.4128}[][][0.65]{$5.4128$}
	 \psfrag{5.413}[][][0.65]{$5.413$}
		 \psfrag{350}[][][0.65]{$350$}
	 \psfrag{400}[][][0.65]{$400$}
	 \psfrag{200}[][][0.65]{$200$}
	 \psfrag{0}[][][0.65]{$0$}
		 \psfrag{Users}[][][0.65]{User}
	 \psfrag{UAV}[][][0.65]{UAV}

		\includegraphics[scale=0.5]{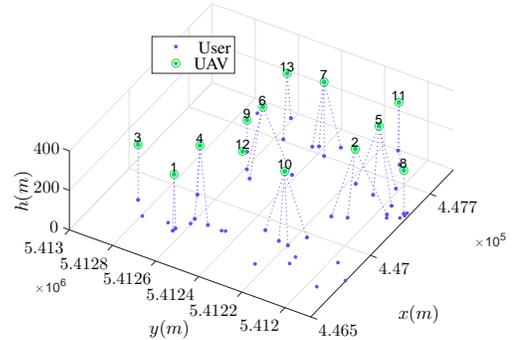}
		\caption{Final 3D placement and UAVs-users association.} 
	\label{AssociationLearn3}
\end{figure}
\vspace{-0.5cm}
\begin{figure}[H]
	\centering
		\psfrag{Final height}[][][0.65]{Final Height}
		\psfrag{x (m)}[][][0.7]{$x (m)$}
		\psfrag{y (m)}[][][0.7]{$h (m)$}
	 \psfrag{350}[][][0.65]{$350$}
	 \psfrag{300}[][][0.65]{$300$}
	 \psfrag{250}[][][0.65]{$250$}
	 \psfrag{200}[][][0.65]{$200$}
	 \psfrag{150}[][][0.65]{$150$}
   \psfrag{100}[][][0.65]{$100$}
	 \psfrag{50}[][][0.65]{$50$}
	 \psfrag{0}[][][0.65]{$0$}
   \psfrag{4.466}[][][0.65]{$4.466$}
	 \psfrag{4.468}[][][0.65]{$4.468$}
	 \psfrag{4.47}[][][0.65]{$4.47$}
	 \psfrag{4.472}[][][0.65]{$4.472$}
   \psfrag{4.474}[][][0.65]{$4.474$}
	 \psfrag{4.476}[][][0.65]{$4.476$}
	 \psfrag{4.478}[][][0.65]{$4.478$}

		\includegraphics[width=8cm,height=5cm]{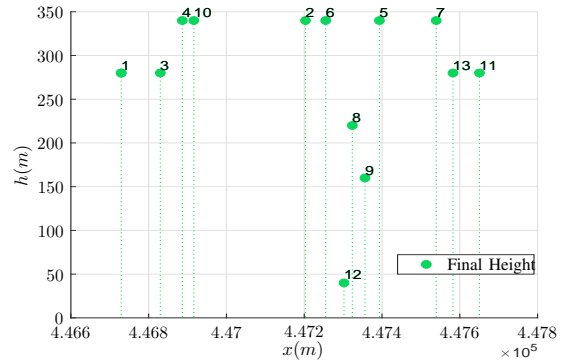}
		\caption{Final heights of UAVs.}
	\label{Height}
\end{figure}
	\vspace{-0.5cm}

\begin{figure}[H]
	\centering
	 \psfrag{Fully Distributed Algorithm}[][][0.7]{Proposed Approach}
   \psfrag{Nearest}[][][0.68]{\textcolor[rgb]{1,1,1}{$....................$}Closest Association}
	 \psfrag{Sum-rate (mbps)}[][][0.7]{Sum-rate (Mbps)}
	 \psfrag{Number of iterations}[][][0.7]{Number of Iterations}
	 \psfrag{3000}[][][0.65]{$3000$}
	 \psfrag{2500}[][][0.65]{$2500$}
	 \psfrag{2000}[][][0.65]{$2000$}
	 \psfrag{1500}[][][0.65]{$1500$}
   \psfrag{1000}[][][0.65]{$1000$}
	 \psfrag{500}[][][0.65]{$500$}
	 \psfrag{0}[][][0.65]{$0$}
   \psfrag{20}[][][0.65]{$20$}
	 \psfrag{40}[][][0.65]{$40$}
	 \psfrag{60}[][][0.65]{$60$}
	 \psfrag{80}[][][0.65]{$80$}
   \psfrag{100}[][][0.65]{$100$}
	 \psfrag{120}[][][0.65]{$120$}
		\includegraphics[width=8cm,height=5cm]{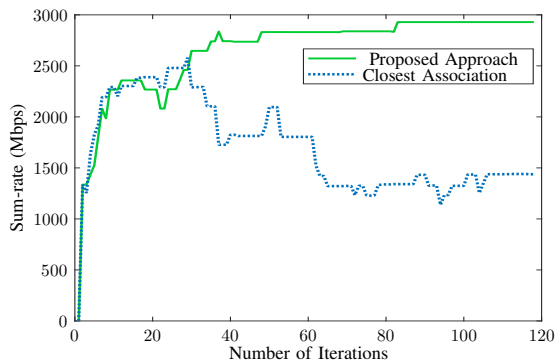}
		\caption{Sum rate convergence.}
	  \label{FinalAssociation}
\end{figure}

Fig.~\ref{AssociationLearn3} shows the final 3D UAVs placement and users association. As expected, the UAVs stand at the center of the served users clusters. Here, it is worth mentioning that the served users are obtained using \textbf{Algorithm}~\ref{MatchingAlgo}. Notice that when an ABS has only one user to serve, it simply stands above the served user in order to improve the LoS probability and enhance the quality of the link. This is for example the case of ABS 3. 
\balance
The final heights of ABSs are better shown in Fig.~\ref{Height} where these altitudes are plotted vs UAVs x-coordinates. It can be seen from the figure that UAVs adjust their heights in order to reduce interferences.  For example, one can remark from Fig.~\ref{AssociationLearn3} that ABSs 6, 9 and 13 are neighbors. On the other hand, Fig.~\ref{Height} shows that each ABS has converged to a different height value which would reduce interferences and, thus, improve the performance over this neighborhood.
  
	Fig.~\ref{FinalAssociation} plots the convergence of the proposed approach vs the number of iterations. The figure shows how the sum-rate evolves over iterations under our proposed scheme that adopts a UAVs-users matching association (as described by Algorithm~\ref{MatchingAlgo}) and under the trivial case where users are connected, at each iteration, to the closest ABS. Clearly, the proposed approach significantly improves the overall sum-rate. For the studied scenario, the sum-rate is improved by 200\% compared with the nearest UAV association. It is worth mentioning that even high performance MINLP solvers may not guarantee the convergence to the global optimum and may only halt at a local optimum. It is also important to note that for each 2D configuration of the network and fixed association matrix, $|\mathcal{H}|^{|\mathcal{B}^A|}=6^{13}$ possible altitudes vectors are to test in order to find the best solution for that configuration. 

	%

\section{Conclusion}
In this paper, we have studied the joint 3D placement and UAVs-users association in UAVs-assisted networks. We have proposed a 3 steps approach that iteratively reaches an efficient solution to the studied optimization problem in only a few number of iterations. In particular, the initial problem was broken into 3 subproblems: UAVs-users associations, 2D positioning of UAVs, and altitudes optimization. Each subproblem has been solved locally using a low-complexity algorithm. Our simulation results have shown appreciable performance of the proposed approach as compared with the trivial case where users are associated, over iterations, to the closest UAV. In ongoing works, we will introduce more uncertainty to the system model and propose a robust approach that takes into account the dynamic nature of the network environment.   


\bibliographystyle{ieeebib}
\bibliography{BiblioUAV}

\end{document}